# Language modulates vision:
# Evidence from neural networks and human brain-lesion models

Author list: Haoyang Chen[1*], Bo Liu[1,2,3,4*], Shuyue Wang[1], Xiaosha Wang[1], Wenjuan Han[5], Yixing Zhu[6,7#], Xiaochun Wang[2,3,4#], Yanchao Bi[1,6,8,9,10, 11 #]

Affiliations:
    1 State Key Laboratory of Cognitive Neuroscience and Learning & IDG/McGovern Institute for Brain Research, Beijing Normal University, Beijing, China, 100875
    2 Department of Radiology, First Hospital of Shanxi Medical University, Taiyuan, Shanxi Province, China, 030001
    3 College of Medical Imaging, Shanxi Medical University, Taiyuan, Shanxi Province, China, 030001
    4 Shanxi Key Laboratory of Intelligent Imaging, First Hospital of Shanxi Medical University, Taiyuan, Shanxi Province, China, 030001
    5 School of Computer Science & Technology, Beijing Jiaotong University, Beijing 100044, China
    6 Institute for Artificial Intelligence, Peking University, Beijing 100871, China
    7 State Key Laboratory of General Artificial Intelligence, Peking University, Beijing 100871, China
    8 School of Psychological and Cognitive Sciences and Beijing Key Laboratory of Behavior and Mental Health, Peking University, Beijing 100871, China
    9 IDG/McGovern Institute for Brain Research, Peking University, Beijing 100871, China
    10 Chinese Institute for Brain Research, Beijing 102206, China
    11 Key Laboratory of Machine Perception (Ministry of Education), Peking University, Beijing 100871, China

* Co-first authors
# Correspondence to: Yixin Zhu (email: yixin.zhu@pku.edu.cn); Xiaochun Wang (email: wangxch@sxmu.edu.cn); Yanchao Bi (email: ybi@pku.edu.cn)

Funding: STI2030-Major Project 2021ZD0204100 (2021ZD0204104 to Y.B.); National Natural Science Foundation of China (31925020 and 82021004 to Y.B.; 32171052 to X.W.; 62376031 to Y. Z.; 62406020 to W. H.); the Fundamental Research Funds for the Central Universities (Y.B.); the PKU-BingJi Joint Laboratory for Artificial Intelligence (Y. Z.)

Acknowledgement: The authors would like to thank Huichao Yang, Ziyi Xiong, Ze Fu, and Haojie Wen for their valuable comments on earlier drafts of the manuscript.

**Abstract**

Comparing information structures in between deep neural networks (DNNs) and the human brain has become a key method for exploring their similarities and differences. Recent research has shown better alignment of vision-language DNN models, such as CLIP, with the activity of the human ventral occipitotemporal cortex (VOTC) than earlier vision models, supporting the idea that language modulates human visual perception. However, interpreting the results from such comparisons is inherently limited due to the "black box" nature of DNNs. To address this, we combined model–brain fitness analyses with human brain lesion data to examine how disrupting the communication pathway between the visual and language systems causally affects the ability of vision–language DNNs to explain the activity of the VOTC. Across four diverse datasets, CLIP consistently outperformed both label-supervised (ResNet) and unsupervised (MoCo) models in predicting VOTC activity. This advantage was left-lateralized, aligning with the human language network. Analyses of the data of 33 stroke patients revealed that reduced white matter integrity between the VOTC and the language region in the left angular gyrus was correlated with decreased CLIP performance and increased MoCo performance, indicating a dynamic influence of language processing on the activity of the VOTC. These findings support the integration of language modulation in neurocognitive models of human vision, reinforcing concepts from vision–language DNN models. The sensitivity of model–brain similarity to specific brain lesions demonstrates that leveraging manipulation of the human brain is a promising framework for evaluating and developing brain-like computer models.

**Introduction**

Recent advancements in artificial neural networks have shown unprecedented capabilities in capturing the responses of the human brain, especially in the field of vision (Bao et al., 2020; Schrimpf et al., 2018; Schrimpf et al., 2021; Yamins et al., 2014;). Image classification models based on deep neural networks (DNNs), such as AlexNet and ResNet (Krizhevsky et al., 2012; He et al., 2016), trained on large image datasets have yielded object representations that are significantly associated with those in the human and macaque ventral occipitotemporal cortex (VOTC), a region crucial for visual object perception and recognition (Kriegeskorte et al., 2008a). Such findings have inspired a line of model–brain fitness analyses that assessed the similarities between the performance of different models (in terms of architecture, parameters, goals, and datasets) and the activity of the VOTC to unravel the representation of the VOTC and assess the biological fitness of DNNs (Dobs et al., 2022; Konkle & Alvarez, 2022; Vinken et al., 2023; Prince et al., 2024).

Recently, multimodal vision models, particularly those trained by aligning image features with language captions, have demonstrated improved performance in brain fitting, brain decoding, and brain-guided image synthesis (Wang et al., 2023; Zhou et al., 2024; Doerig et al., 2022; Conwell et al., 2023; Oota et al., 2022; Luo et al., 2023; Luo et al., 2024). These findings provide evidence of the influence of language on the visual cortex, a classic topic that remains highly controversial in the field of cognitive neuroscience (see Lupyan, 2016; Thierry, 2016). Supporting cognitive evidence comes from prominent behavioural cross-linguistic studies and verbal learning experiments, which have demonstrated that verbal labels can influence visual colour categorization (e.g., Gilbert et al., 2006; Drivonikou et al., 2007; Winawer et al., 2007). However, the robustness of these cognitive findings has been challenged (e.g., Martinovic et al., 2020; see SI of Fedorenko et al., 2024), and even when such effects are actually observed, researchers are unable to determine whether they arise from the perceptual stages or other cognitive processes (Maier & Abdel Rahman, 2019).

The results from the new vision-language model-brain fitness approach, while compelling, are inconclusive for several reasons. A prominent model used in these studies is Contrastive Language-Image Pretraining (CLIP), the extensiveness of the dataset on which it was trained potentially contributing substantially to its performance (Conwell et al., 2024a). Wang et al. (2023) compared the performance of a CLIP-style model with an unsupervised model trained on the same dataset (i.e., YFCC15M) and reported that, after controlling for the details of the training dataset, the CLIP-style model showed diminished performance in fitting to the activity of the VOTC. The remaining effects were primarily localized to the bilateral extrastriate body area and fusiform face area, potentially reflecting specific influences regarding human interactions within the images (see discussion in Wang et al., 2023). In addition, the empirical robustness of these findings requires further validation, as current comparisons predominantly rely on a single fMRI dataset (*Natural Scenes Dataset*, Allen et al., 2022).

More critically, the model-brain fitness approach suffers from an inherent challenge in interpretation because of the "black box" nature of the learned representations of the DNN. Even when the CLIP model differ from control models solely by the addition of language supervision, their superior explanatory power for the activity of brain regions might stem from their ability to capture higher-order object-relational structures that are not attributable only to language, since language encodes relational structures that overlap with a range of nonlinguistic high-level relations (e.g., "hammer" and "hand" are similar in both linguistic and visual-action association spaces). This alternative hypothesis is supported by a study demonstrating that pure-language models perform comparably to vision models in predicting activity in both the macaque inferior temporal cortex and the human VOTC (Conwell et al., 2024b). Therefore, more direct testing is needed to determine whether the unique effects of language–vision models genuinely reflect how the language system modulates visual perception in the human brain.

To address these issues, we investigated the influence of language experience on activity in the VOTC in response to object images through a combination of human brain lesion data, fMRI data, and computational modelling approaches. Study 1 evaluates whether, compared with unsupervised vision models, vision models aligned with different scales of language information (e.g., sentence-level vs. word-level alignment) consistently provide unique explanatory power for the patterns of activity of the human VOTC across four datasets featuring diverse picture stimuli, language experiences, and tasks (Fig. 1). Study 2 provides a crucial test of causality through human brain-lesion models (Fig. 4). Specifically, in a group of patients with brain damage (N = 33), we examined whether reducing the structural integrity of the connections between the VOTC and language regions affects the ability of different computational models to explain the activity of the VOTC during visual processing.

**Results**

**Study 1: Computational models fitting four brain datasets from healthy populations with different language experiences**

To investigate whether and how language supervision influences the explanatory power of the representations within the VOTC, we compared three computational vision models with different degrees of language supervision with representational similarity analysis (RSA; Kriegeskorte, Mur, & Bandettini, 2008b) across four distinct fMRI datasets. These datasets are characterized by different participant populations, tasks, and language experiences, allowing us to examine the generalizability of the effects of language on visual perception.

Object neural representation in the four fMRI picture viewing datasets

We included four fMRI datasets from healthy subjects viewing object pictures, encompassing different object types and tasks. The first three datasets were derived in-house (see Methods), and the fourth was an open dataset. Object neural activity patterns in the VOTC are included in each fMRI dataset and used to construct the object representation dissimilarity matrices (neural RDMs).

*OPN95 (oral picture naming) & SPN95 (sign language picture naming).* The participants viewed colour pictures of 95 objects across three categories (32 animals, 35 small, manipulable objects, and 28 large, nonmanipulable objects) and were instructed to name them. The *OPN95* dataset includes the data from typically developed participants (i.e., those with speech experience; N=26 in the analyses) performing oral picture naming, whereas the *SPN95* dataset includes the data from early deaf individuals (i.e., those with sign language experience; N=32 in the analyses) performing sign language naming with their nondominant hand. The data from these two groups were analysed both separately and collectively.

*FV14 (colour knowledge retrieval).* The participants (N=33) viewed grayscale pictures of 14 fruits/vegetables (one image per object) and were asked to judge whether the typical skin colour of the object was red.

*THINGS.* An open dataset (Hebart et al., 2023) in which participants (N=3) viewed 720 objects from various categories (e.g., animals, small, manipulable objects, large, nonmanipulable objects, plants, fruits), with each category containing 12 unique images. The participants were instructed to maintain fixation on the centre of the screen and respond to occasional artificially generated images.

Object model representation in three computation vision models

We considered three computer vision models with different levels of language involvement during pretraining (Fig. 1): 1) CLIPvision: The visual encoder of CLIP, supervised by natural language (sentence descriptions). CLIP is trained with image/caption pairs to align features between the images and the text, and sentence-level descriptions containing verbal labels and object relationships are integrated into the visual encoder of CLIP using a multimodal loss signal in the final layer. 2) ResNet-50: A supervised image classification model trained to predict human-generated verbal labels (word-level categorization) for input images. ResNet-50 uses image-label pairs to minimize the classification error. 3) MoCo: A self-supervised visual representation learning model trained exclusively on images. The distance between an image and its transformed version is minimized, while the distance between different images is maximized.

The identical architectures of CLIPvision, ResNet-50, and MoCo enabled direct comparisons of the object set representations in the corresponding layers. Image features were extracted from the three models spanning 18 matched layers, ranging from early to late stages. Model RDMs were generated by computing the Pearson distances (1-Pearson's correlation) between feature vectors for each object pair. For the THINGS dataset, which includes multiple images per concept, the RDMs were calculated using the average features of the images corresponding to the same concept. The final pooling layer, which showed the lowest average intermodel correlation (Fig. 2a)—indicating more distinct and specialized representations across models—was selected for subsequent brain-fitting analyses. The resulting RDMs of the final pooling layer illustrated moderate intercorrelations across the *OPN95/SPN95*, *FV14*, and *THINGS* datasets, with Spearman's correlation coefficients ranging from 0.43 to 0.63 (Bonferroni-corrected $p < 0.05$). A representative visualization of the RDMs from the *THINGS* dataset is provided in the left panel of Fig. 2b.

To assess similarity between the model and human behavioural object-relatedness, we conducted correlation analyses between the model-generated and the human behavioural RDMs. The human behavioural RDMs were generated by averaging ratings across participants within distinct groups. For the *OPN95/SPN95* and *FV14* datasets, object-semantic similarity was evaluated through a 7-point Likert scale (1: least similar; 7: most similar). For the *THINGS* dataset, object similarity was assessed with an odd-one-out task, and the probability of participants selecting objects i and j as belonging together was quantified (Hebart et al., 2023).

The results are summarized in Fig. 2c, where each bar represents one model (CLIP, ResNet, MoCo) and contains three data points (corresponding to the three datasets). As the degree of language involvement increased (from MoCo to ResNet and to CLIP), the alignment with human behaviour progressively improved, demonstrating the relative advantage of language information alignment in capturing semantic relationships. Partial correlation analyses examining each model's unique contribution to predicting human behaviour beyond what is shared with the other two models likewise revealed a progressive increase in behaviour alignment from MoCo to ResNet to CLIP, reinforcing the benefits of language supervision. The right panel of Fig. 2b highlights the top 2 image pairs in each model's RDM within the *THINGS* dataset after partialing out the other two models. These examples illustrate the heavier emphasis of CLIP on semantic properties, the moderate semantic focus of ResNet, and the predominant reliance of MoCo on lower-level visual attributes, thus revealing a continuum of semantic–visual encoding across the three models.

Object model representations fitting brain activity representations in the VOTC: RSA results

We next examined how the model representations relate to neural data with searchlight RSA in the VOTC mask, defined by contrasting pictures with baseline in the *OPN95* dataset (FDR q < 0.05; see Fig. S1 & Methods for details). For each participant, the Spearman correlation was assessed between the model visual RDMs and the neural RDMs within a 10 mm radius searchlight sphere. We focused on testing whether CLIPvision and ResNet provided additional explanatory power over MoCo to assess the effects of sentence description and verbal categorization, respectively.

Specifically, to evaluate the relative specific contribution of the alignment with *sentence descriptions* (i.e., the introduction of word relations), we computed the partial Spearman correlation coefficient between the CLIPvision RDM and the neural RDMs, with the ResNet RDM and MoCo RDM serving as covariates. To evaluate the specific relative contribution of the alignment with *verbal categorizations*, we computed the partial Spearman correlation coefficient between the ResNet RDM and the neural RDMs, with the MoCo RDM serving as a covariate. Below, the results are shown at a statistical voxel-level threshold $p < 0.001$, one-tailed, cluster-level familywise error (FWE)- corrected $p < 0.05$, unless otherwise specified.

*Sentence description effect (CLIPvision over ResNet and MoCo)*. In the *OPN95* dataset, group-level one-sample t tests on the Fisher z-transformed partial Spearman correlation coefficients for all participants revealed two significant clusters: one in the left lateral occipital complex (L-LO) and another in the left fusiform gyrus (L-FG) extending into the left inferior temporal gyrus (L-ITG) (Fig. 3, top left panel, blue). In the *SPN95* dataset, similar significant clusters were found, but with more medial peak locations (Fig. 3, top middle panel, blue). These effects were not affected by auditory experience (hearing vs. deaf) or the language modality (oral vs. sign language naming): region of interest (ROI) analysis of these two clusters revealed statistically comparable effects in the deaf and hearing groups (mean difference ± SE = 0.007 ± 0.010, two-tailed $p = 0.508$; $BF_{10} = 0.322$, moderate evidence in support of H0: deaf = hearing). In the *FV14* dataset, analyses revealed a significant cluster in the left occipital pole (L-OP) (Fig. 3, top right panel, blue). In the *THINGS* dataset, individual-level analysis was conducted for each subject at a voxel-level $p < 0.001$, one-tailed, cluster-level FWE-corrected $p < 0.05$. Across all the subjects, the analysis revealed significant clusters in the bilateral lateral occipital complex, extending to the bilateral lingual gyrus, fusiform gyrus, parahippocampal gyrus (PHG), and inferior temporal gyrus (Fig. 3, bottom panel, blue).

*Verbal categorization effect (ResNet over MoCo).* In the *OPN95* dataset, group-level, one-sample t tests on the Fisher z-transformed partial Spearman correlation coefficients for all participants revealed significant clusters in the bilateral occipital pole (OP), both extending to the lingual gyrus (LING), the lateral occipital complex, and the posterior fusiform gyrus (Fig. 3, top left panel, orange). In the *SPN95* dataset, the analyses revealed similar clusters. ROI analysis of these two clusters revealed no significant difference between the two groups (hearing vs. deaf: mean difference ± SE = 0.002 ± 0.008, two-tailed $p = 0.764$; $BF_{10} = 0.278$, moderate evidence in support of H0: deaf = hearing). In the *FV14* dataset, no cluster survived at the conventional threshold (voxel-level $p < 0.001$, one-tailed, cluster-level FWE-corrected $p < 0.05$). In the *THINGS* dataset, across all subjects, the analysis revealed significant clusters in the bilateral lingual gyrus, extending to the occipital pole, occipital fusiform gyrus, lateral occipital complex, and intracalcarine cortex (Fig. 3, bottom panel, orange).

RSA-based lateralization results

Consistent with the well-established left lateralization of language networks in the human brain (see review in Güntürkün et al., 2020), the above searchlight results of the sentence description effect revealed larger, positive clusters on the left VOTC than on the right VOTC. In this section, we quantified and examined these potential VOTC laterality effects. We computed the laterality index (LI) in the VOTC for each dataset with the LI toolbox (Wilke & Lidzba, 2007) implemented in SPM12 (for details, see Methods). At the group level, the LIs were computed by extracting t values generated from the one-sample t tests against zero conducted above for RSA within the whole VOTC mask. At the individual level, the RSA-derived Fisher z-transformed rho values within the VOTC mask were used to compute the LIs, followed by one-sample t tests against zero. In the *THINGS* dataset, individual subject LIs were calculated following the same procedure but without statistical testing. The group-level LI ranges from -1 to 1, where -1 indicates complete right lateralization, +1 indicates complete left lateralization, and values between -0.2 and 0.2 are considered to indicate bilateralization (Seghier, 2008).

*Sentence description effect.* We observed a significant left-lateralization trend for the sentence description effect in the VOTC at both the group and individual levels. At the group level, the *OPN95*, *SPN95*, and *FV14* datasets all showed strong left lateralization (*OPN95*: 0.64; *SPN95*: 0.68; *FV14*: 0.68). At the individual level, as shown in Fig. 3 (blue bar), the results revealed a significant left-lateralized sentence description effect for the three datasets (*OPN95*: mean LI ± SE = 0.175 ± 0.071, two-tailed $p < 0.05$, Cohen's *d* = 0.480; *SPN95*: mean LI ± SE = 0.339 ± 0.066, two-tailed $p < 0.001$, Cohen's *d* = 0.926; *FV14*: mean LI ± SE = 0.348 ± 0.072, two-tailed $p < 0.001$, Cohen's *d* = 0.846). In the *THINGS* dataset, all the subjects demonstrated stable and pronounced left-lateralization (Sub01: 0.60; Sub02:

0.52; Sub03: 0.45).

*Verbal categorization* effect. The verbal categorization effect exhibited no clear lateralization pattern at either the group or individual level. At the group level, the results for the *OPN95*, *SPN95*, and *FV14* datasets indicated an evenly distributed pattern within the VOTC (*OPN95*: 0.04; *SPN95*: 0.03; *FV14*: -0.05). At the individual level, as shown in Fig. 3 (orange bar), the LIs for these datasets remained relatively balanced, with no statistically significant lateralization detected (*OPN95*: mean LI ± SE = 0.083 ± 0.062, two-tailed $p$ = 0.097, Cohen's $d$ = 0.262; *SPN95*: mean LI ± SE = 0.061 ± 0.055, two-tailed $p$ = 0.138, Cohen's $d$ = 0.196; *FV14*: mean LI ± SE = -0.024 ± 0.076, two-tailed $p$ = 0.622, Cohen's $d$ = 0.055). Similarly, in the *THINGS* dataset, the three subjects displayed no stable lateralization pattern (Sub01: 0.43; Sub02: -0.31; Sub03: 0.46).

Note that the LI for the raw effects of the CLIPvision model prior to the partial correlation analyses was not stable across the three datasets, suggesting that the strong left lateralization in the *sentence description* effects was attributable to the CLIPvision representation, which is relatively specific beyond ResNet and MoCo.

**Study 2: Causal testing of model-fitting-brain effects in patients with brain damage**

In Study 1, we observed the specific relative advantages of CLIPvision over ResNet and MoCo in explaining left VOTC responses to object images across four datasets. In Study 2, we causally tested whether such advantages were indeed related to vision-language alignment with brain lesion data. Specifically, we examined whether these effects were modulated by the integrity of the communication pathway between the left VOTC and the higher-order language system. We reasoned that if the CLIPvision model's effect diminished following damage to vision-language communication centres, it would provide positive evidence for the language-alignment origin of the CLIPvision effect advantage regarding the representations within the left VOTC. We tested this in a group of patients with varying degrees of brain damage in terms of lesions affecting the white matter (WM) structural connections between the language network and VOTC while sparing the ventral visual cortex (in-house data collected for another project investigating colour representations; see Methods for details).

Patient picture-viewing fMRI data and white matter structural data.

A group of patients with brain damage (N = 33) underwent the same fMRI scans as did the healthy controls (N = 33) in the *FV14* dataset described above. In addition to fMRI data, we collected structural and high angular resolution diffusion imaging (HARDI) data from both the patient and healthy control groups to assess the integrity of the WM connections of interest. We investigated the associations between lower integrity in the language–vision system WM connections and different vision–model–RSA effects. The distribution of the lesions in the 33 patients is shown in Fig. 4, revealing a typical middle cerebral artery stroke pattern, with lesions widely distributed across grey matter (GM) regions and WM tracts.

To evaluate the integrity of the left vision-language WM connections in the patients, we first mapped the WM connections by establishing a WM mask of interest with the HARDI data in the healthy control group. We performed tractography between the left VOTC mask and the cortical language regions defined by a commonly used language mask (Fedorenko et al., 2010, contrasting intact sentences to nonword lists in 220 subjects; see Fig. 4 "Language Region"). The mask encompasses language-activated clusters located in the left angular gyrus (L-AG), left posterior middle temporal gyrus (L-pMTG), left dorsolateral anterior temporal lobe (L-dlATL), left inferior frontal gyrus (L-IFG), left inferior frontal gyrus orbital part (L-IFGorb), and left middle frontal gyrus (L-MFG). The WM connections between the left VOTC mask and each language region, as well as the left VOTC mask connecting all six regions together, were traced and binarized. We applied an individual-level probability threshold of 0.1, a group-level threshold of 0.5, and an explicit WM mask probability threshold of 0.4. For each stroke patient, the mean fractional anisotropy (FA) values within each WM mask were computed to quantify the integrity of the left vision-language

WM connections.

Association between left vision-LAG WM integrity and sentence description effects on the left VOTC

Following a similar RSA procedure to that used in Study 1, we quantified how well each of the three vision models (CLIPvision, MoCo, and ResNet) explained the neural activity patterns in the left VOTC during picture viewing for each patient, as well as their specific effects.

We tested whether the integrity of the left vision-language WM connections (beyond total lesion volume) was associated with different vision-model brain-fitness effect patterns via multiple regression analyses, where the patients' FA values of the left vision-language WM connections served as the dependent variable, and the three vision models' brain-fitness measures (Fisher z-transformed RSA Spearman's correlation coefficients within the left VOTC mask) were included as independent variables, along with total lesion volume. As shown in Fig. 5a, the regression model for the connection linking the left VOTC and the L-AG was significant ($F_{4,28}$ = 3.50, $R^2$ = 0.33; $p <$ 0.05; Table 1). That is, the structural integrity of the left vision–AG pathway was significantly associated with how well the vision models fit the neural data. Significant but opposite effects were observed for the CLIPvision model and MoCo model ($p < 0.05$; for details, see Table 1). Greater integrity of the left vision–AG brain connection was associated with better left VOTC activity fitness to the CLIPvision model and weaker fitness to the MoCo model, whereas left vision–AG brain connection disruption was associated with weaker CLIPvision model fitness and stronger MoCo model fitness.

We also validated these results with partial correlation analyses, in which the correlations between each vision model's specific effect on the left VOTC to the left vision–AG WM integrity was assessed while controlling for total lesion volume. As shown in Fig. 5b, these analyses confirmed the regression model findings: left vision–AG integrity was positively correlated with the brain-fitness measure of CLIPvision ($r = 0.41$, $p < 0.01$) and negatively correlated with that of MoCo ($r = -0.50$, $p < 0.005$).

Taken together, these results provide positive evidence that the unique brain-explanatory power of the CLIPvision model, with respect to ResNet and MoCo, is indeed related to its alignment with language information. Damage to brain structural connections between the VOTC and language-activated AG clusters weakens the advantage of the CLIPvision model while increasing the performance of the MoCo model in explaining brain response patterns during object viewing.

*Validation: No effects of the right homologous vision–AG WM connection.* To confirm that the observed modulatory effects arose specifically from the language-related functions of the left AG rather than its nonlinguistic (e.g., multimodal integration) capabilities, we examined the analogous white matter connections in the right hemisphere. The right AG is also involved in multimodal integration but lacks the same degree of language specialization as the left AG. Therefore, if the observed effects were solely due to the nonlinguistic, bilateral functions of the AG, we would expect to encounter similar correlations in the right hemisphere. No such effects were observed: for the CLIPvision-specific effect, $r = -0.12$ (one-tailed $p = 0.735$) in the left VOTC, and for the MoCo-specific effect, $r = 0.12$ (one-tailed $p = 0.741$) in the left VOTC (Fig. 5c). Similarly, no significant effects were found for the right VOTC-right AG tract: CLIPvision-specific effect, $r = -0.04$ (one-tailed $p = 0.590$) in the right VOTC; MoCo-specific effect, $r = 0.02$ (one-tailed $p = 0.549$). These null results in the right hemisphere support the interpretation that the modulation stems from the language-related properties of the left AG rather than from general multimodal integration functions.

**Discussion**

Whether human visual cortex activity during visual perception is modulated by language experience remains debated. Our results show that vision DNN models with language alignment supervision—both sentence-level (CLIPvision) and category-label-level (ResNet)—consistently demonstrated unique explanatory power for the activity of the human VOTC across four datasets with respect to an unsupervised model (MoCo). The sentence-level advantage exhibited by CLIPvision demonstrated left-hemisphere lateralization, which aligns with the lateralization of the language network. Analyses of the data of 33 stroke patients indicated that the responses in the left VOTC to DNN models depended on the structural integrity of the left vision–angular gyrus WM connection. Reduced tract integrity weakened the CLIPvision-fitting effect while enhancing the MoCo-fitting effect, independent of the degree of overall brain damage.

Recent research comparing CLIP and ResNet in fitting to the responses in the human VOTC with the NSD dataset suggested the presence of CLIP-specific effects, although the robustness and interpretation of their results are controversial (Wang et al., 2023; Conwell et al., 2024a). Our findings validate the robustness of these effects across four datasets of differing stimulus types (isolated objects without backgrounds: *FV14*, *OPN95*/*SPN95*; natural images: *THINGS*), populations (healthy typical populations with speech experience, early-deaf individuals with sign language experience), and task demands (image oddball detection, oral picture naming, sign language picture naming, and colour knowledge retrieval). CLIP consistently demonstrated unique explanatory power for the activity of the VOTC across all four datasets with respect to the label-supervised (ResNet) and unsupervised (MoCo) vision models.

Critically, two pieces of evidence suggest that the advantage of CLIP in fitting the neural responses within the VOTC stems from the language system in the human brain rather than from the ability to capture some kind of nonverbal, higher-order relations. First, the fitting advantage of CLIP demonstrated consistent leftward lateralization across all datasets, with stronger effects in the left VOTC than in the right VOTC. This aligns with the left-lateralized characteristic of the human language network, which is not reflected as stably in any other system (Corballis, 2012). Second, in stroke patients, the compromised integrity of the WM tract connecting the left VOTC and the AG (language cluster) diminished the ability of CLIP and enhanced the ability of MoCo to fit the neural activity within the VOTC. This finding suggests that VOTC activity reflects the representations of CLIP or MoCo depending on the relevant communication efficiency with the language brain system. Notably, this effect was not attributable to overall lesion severity or general AG functioning, as connections to the right AG showed no modulation effects.

These findings indicates that CLIP's visual-language alignment training likely parallels those in the human brain and is also critical for addressing a classical question in human cognition. In the human ventral visual system, although there is increasing evidence of nonvisual influences on the activity patterns of the VOTC, the effects of language have not been established. For example, when nonvisual stimuli, such as object names or tactile inputs, are presented—even to congenitally blind individuals—the VOTC exhibits response profiles that are at least partially similar to those during visual perception, particularly in terms of object category preference distributions (Mattioni et al., 2020; van den Hurk et al., 2017; Peelen et al., 2013; Wang et al., 2015; see Ricciardi et al., 2014 & Bi et al., 2016, for reviews). These findings are commonly interpreted as reflecting associations or mappings between visual and nonvisual object properties, such as related actions (Bi et al., 2016; Mahon et al., 2007). Here, we show that incorporating language inputs is important for understanding human VOTC.

How does language shape the activity of the VOTC during visual perception? Our findings provide several key clues to answering this classical question. First, we observed distinct effects across the three vision-language models: with respect to MoCo, CLIPvision showed consistent, positive, and left-lateralized effects, whereas ResNet's effects were more variable, bilateralized, and unrelated to vision-language connection integrity. As explained earlier, these models reflect two potential pathways through which language may influence visual processing: verbal labels facilitate specific object categorization (Gelman & Roberts, 2017), and larger linguistic units introduce broader relational structures to visual concepts (Unger & Fisher, 2021). The robust advantages of CLIP over ResNet suggests a key role for relational structures among word combinations in larger linguistic units.

The second clue stems from the functionality of the left AG, whose connection with the VOTC modulates vision model fitting patterns. The left AG is widely regarded as a cross-modal hub that integrates different semantic networks related to language and multisensory experiences (Xu et al., 2017; Seghier, 2013). It processes both taxonomic and thematic object relations (Xu et al., 2018; Schwartz et al., 2011) and supports semantic composition (Zhang et al., 2022), potentially facilitating interactions between the language system's relational structures and the activity of the VOTC. The absence of significant findings regarding connections between the VOTC and other language regions (e.g., the anterior temporal lobe) should be interpreted with caution, as negative results are inherently difficult to parse. Future studies using different multimodal models, such as those with fusion-encoder (e.g., FLAVA) or encoder-decoder (e.g., MetaLM) architectures, may better illuminate these regions' roles.

A final key clue is the patient-level result showing reduced CLIPvision effects following the disruption of VOTC-language AG connections. This observation challenges the view of language modulation as merely a "pretraining" process that creates fixed object representations within the VOTC. Instead, our results suggest at least two alternative mechanisms. One is an online interactive process, in which the visual parsing of the VOTC communicates with language system representations, enabling real-time modulation of the activity of the VOTC (see Lupyan et al., 2020, for a detailed discussion on online vs. offline effects). The other is a poststroke plasticity process (approximately 3 months in our chronic patients), wherein disconnection of the visual cortex from the language system induces plastic changes in the VOTC, leading to a more visually driven state. Both interpretations emphasize the dynamic nature of language–vision interactions.

Several open questions remain. First, pinpointing the specific anatomical locations corresponding to the effects of CLIPvision across datasets remains challenging owing to the differences in the stimuli and tasks. The VOTC exhibits considerable heterogeneity in object category preferences and visual attribute preferences and its sensitivity to task demands (e.g., Wang et al., 2015). The mechanisms by which language experiences interact with these factors likely involve complex dynamics that require further study. Moreover, converging evidence suggests that language plays a nuanced role in the activity of VOTC during perception, being neither necessary nor sufficient. Many core features of the organisation of the human VOTC are observed in the IT cortices of nonhuman primates despite their lack of developed language systems (Kriegeskorte et al., 2008a; Bao et al., 2020) and have emerged in unsupervised vision DNN models (Prince et al., 2024; Konkle et al., 2022). Additionally, purely language-derived information without sensory input, such as when congenitally blind individuals retrieve object colour knowledge, fails to activate the visual cortex (Wang et al., 2020; reviewed in Bi, 2021). The specific functional and behavioural relevance of language modulation of the human VOTC warrants further investigation.

To conclude, our findings indicate that vision models employing language supervision uniquely capture representations in the left hemisphere of the human visual cortex. The strong dependence of this effect on left VOTC–AG WM tract integrity suggests that the human VOTC actively integrates visual inputs with language experience during perception. These findings indicate that the human visual cortex engages in dynamic computations that align visual inputs with language experiences during perception. The sensitivity of model–brain similarity to brain lesions, which is specifically related to the properties of the model, further highlights the leveraging of human brain manipulation as a promising framework for evaluating and developing brain-like machine models.

**Methods**

Study 1: fMRI dataset collection

Participants

All the participants in the fMRI tasks were right-handed except for two left-handed individuals (ID: D010, N001) and three ambidextrous individuals (ID: D017, D018, D022) in the SPN95. The participants in the *OPN95* and *FV14* datasets were all native Mandarin speakers, the deaf participants in the SPN95 were fluent Chinese Sign Language (CSL) users (11 native signers, others' age of sign language acquisition < 12 years), whereas those in the *THINGS* dataset were all native English speakers. The participants in *SPN95* were all congenital or early-deaf individuals. The sample sizes of the OPN95, SPN95, THINGS, and FV14 databases were 26 (18 females; median age: 20 years; range: 18–32 years), 36 (19 females; median age: 32 years; range: 22–45 years), 3 (2 females; mean age at the beginning of the study: 25 years), and 33 (12 females; median age: 40 years; range: 20–65 years), respectively.

For the behavioural tasks, independent groups were recruited for OPN95/SPN95, FV14, and THINGS. However, some participants in the FV14 dataset overlapped with those in the fMRI task. Specifically, the sample sizes for the behavioural task were as follows: OPN95/SPN95 (100 participants, 71 females; median age: 21 years; range: 18–26), FV14 (36 participants, 14 females; median age: 36 years; range: 20–65), and THINGS (14,025 participants; for details, see Hebart et al., 2023). All participants in the OPN95/SPN95 and FV14 datasets were native Mandarin speakers, whereas those in the THINGS dataset were native English speakers.

*Ethics approval*

All protocols and procedures of the current study were approved by the Ethics Committee of the State Key Laboratory of Cognitive Neuroscience and Learning at Beijing Normal University (ICBIR_A_0040_008), the Ethics Committee of the First Hospital of Shanxi Medical University (No. 2021-K035) and the Human Subject Review Committee at Peking University (2017-09-01). Prior to participation, all participants provided written informed consent. The study was conducted in accordance with the Declaration of Helsinki and adhered to all relevant ethical guidelines.

Stimuli and procedures for the task-fMRI experiment

*OPN95/SPN95.* A total of 95 objects were chosen, encompassing 3 common categories (32 animals, 35 small manipulable objects, and 28 large nonmanipulable objects). Each object includes one sample presented as a 400 × 400 pixel, coloured image with the representative exemplar of the object presented against a white background (10.55° × 10.55° of visual angle). All the participants were asked to name the pictures displayed with oral (hearing) or sign (deaf, with their left hand) language. The whole experiment included 6 runs, with each item being repeated 6 times. Each run (8 min 45 s) consisted of 95 trials, in which each item was presented once. Each trial consisted of 0.5 s of fixation, 0.8 s of stimulus presentation and an intertrial interval (ITI) ranging from 2.7 s-14.7 s. Each run began and ended with a 10 s fixation.

*THINGS.* A total of 720 objects were chosen, including multiple categories such as animals, tools, fruits, and bodies. The participants viewed images of 720 representative object concepts and were instructed to fixate on a central point and press a button upon viewing artificial images. Each concept includes 12 exemplars, i.e., 8640 unique images in total. The whole experiment included a total of 15–16 scanning sessions, among which the first 1–2 sessions were dedicated to testing the reliability of the head fixation model and obtaining functional localizers and retinotopic maps. The subsequent 12 runs consisted of a unique sample of the aforementioned 720 concepts, in which each item was presented once. All the presented images subtended 10 degrees of visual angle and were presented on a grey background for 0.5 s and overlaid with a fixation crosshair subtending 0.5 degrees, followed by a 4-s rest stage.

*FV14.* Fourteen fruits/vegetables were chosen. Each object included one sample presented as a 400 × 400 pixel, grayscale image with the representative exemplar of the object presented against a white background (6.22° × 6.22° of visual angle). The experiment included 4 runs, each of which consisted of 32 1 s-long stimulus trials and 32 11 s-long null trials. Each image was presented twice within each run. The order of the 32 trials was pseudorandomized while ensuring that no two consecutive trials were identical. Each run began with a 12-s silence period and ended with a 4-s silence period.

Image acquisition

   *OPN95/SPN95.* Functional and anatomical MR images were collected at the MRI centre of Beijing Normal University with a 3-Tesla Siemens Trio Tim scanner. High-resolution 3D structural images were collected with a 3D magnetisation prepared-rapid gradient echo (MPRAGE) sequence in the sagittal plane (144 slices, TR = 2530 ms, TE = 3.39 ms, flip angle = 7°, matrix size = 256 × 256, and voxel size = 1.33 × 1 × 1.33 mm). Functional images were acquired with an echo–planar imaging (EPI) sequence (33 axial slices, TR = 2000 ms, TE = 30 ms, flip angle = 90°, matrix size = 64 × 64, and voxel size = 3 × 3 × 3.5 mm with a gap of 0.7 mm).

   *THINGS.* Functional and anatomical MR images were collected at the National Institutes of Health (NIH) in Bethesda, MD (USA) with a 3 Tesla Siemens Magnetom Prisma scanner and a 32-channel head coil. High-resolution 3D structural images were collected with an MPRAGE sequence (208 sagittal slices, voxel size = 0.8 × 0.8 × 0.8 mm, TR = 2.4 s, TE = 2.24 ms, matrix size = 320 × 300, FOV = 256 × 40 mm, flip angle = 8°). The whole-brain functional MR data were collected in 2 mm isotropic resolution (60 axial slices, voxel size = 2 × 2 × 2 mm, TR = 1.5 s, TE = 33 ms, matrix size = 96×96, FOV = 192 × 192 mm, flip angle = 75°).

   *FV14.* Functional and anatomical MR images were collected at the Department of Magnetic Resonance Imaging, First Hospital of Shanxi Medical University, with a 3T Siemens Magnetom Skyra scanner. High-resolution 3D structural images were acquired with an MPRAGE sequence (sagittal slices, TR = 2530 ms, TE = 2.88 ms, flip angle = 7°, matrix size = 224 × 256, interpolated to 448 × 512, voxel size = 0.5 × 0.5 × 1 mm³, FOV = 224 × 256 mm²). Functional images were acquired with a multiband EPI sequence (axial slices, TR = 2000 ms, TE = 30 ms, flip angle = 90°, matrix size = 72 × 72, voxel size = 2.5 × 2.5 × 2.5 mm³, FOV = 180 mm × 180 mm, multiband factor = 2).

Preprocessing for the task-fMRI data

   *OPN95/SPN95.* The functional images were preprocessed and analysed with Statistical Parametric Mapping (SPM12; http://www.fil.ion.ucl.ac.uk/spm). For each participant, the first five volumes of each run were discarded to ensure signal equilibration. The remaining images were subsequently corrected for slice timing and head motion and then spatially normalized to the Montreal Neurological Institute (MNI) space via unified segmentation (resampled to a 3 × 3 × 3 mm³ voxel size). The data of three hearing participants and four deaf participants were excluded from the analyses because of excessive head motion (> 3 mm/3°). The object-relevant beta weights of the functional images of each participant were obtained with a general linear model (GLM) that contained an onset regressor for each of 95 items, 6 regressors of no interest corresponding to the 6 head motion parameters, and a constant regressor for each run. Each item-relevant regressor was convolved with a canonical haemodynamic response function (HRF), and the high-pass filter cut-off was set as 128 s. The resulting *t*-maps for each item with respect to the baseline were used to create neural RDMs.

   *THINGS.* The preprocessing pipeline for the functional images included slice-timing correction, rigid-body head-motion correction, susceptibility-distortion correction on the basis of the field maps, spatial alignment to each participant's T1-weighted anatomical reference images, and brain tissue segmentation and reconstruction of the pial and WM surfaces. The single-trial beta weights were estimated with a GLM. The mean beta-maps of 12 samples for each concept were used to create neural RDMs.

   *FV14.* The functional images were preprocessed and analysed with SPM12 following the same procedure used for the OPN95/SPN95 databases. For each participant, the first six volumes of each run were discarded for signal equilibration; the remaining images were subsequently corrected for time slicing and head motion and then spatially normalized to the MNI space via unified segmentation (resampling into a 2 × 2 × 2 mm³ voxel size). One or two runs from four subjects showed excessive head motion (> 2.5 mm/2.5°) and were excluded from the analysis. The single-trial beta weights were estimated with a GLM, after which the t-maps for each item versus baseline were estimated and used to create neural RDMs.

*Behavioural object relatedness rating*

For the OPN95/SPN95 and FV14 datasets, the participants were asked to rate the similarity of the same objects presented during the fMRI task pairwise on a 7-point Likert scale (1 = least similar, 7 = most similar). For the THINGS dataset, participants were presented with ternary object triplets and asked to click on the object they perceived as the "odd-one-out" in each triplet. Not all triplets were evaluated by every participant, resulting in a total of 5,517,400 triplet choices (for further details, see Hebart et al., 2023).

Study 1: Analysis procedures

Model details and feature extraction

The models used in the analyses included (1) CLIP trained by OpenAI (with a ResNet-50 backbone) (Radford et al., 2021); (2) ResNet-50 (He et al., 2016) pretrained on ImageNet-1k; and (3) MoCo v3 pretrained on ImageNet-1k (with a ResNet-50 backbone) (Chen et al., 2021).

Each image was preprocessed according to the preprocessing parameters provided by the pretrained model and passed through the model to extract the outputs from each layer. As the differences in features among the models were most pronounced at the penultimate layer (see Fig. 2), we selected the visual encoder output in CLIP and the penultimate layer outputs of ResNet-50 and MoCo v3 for fitting the neural representations. As none of the models performed well in recognizing or classifying grayscale images, we passed the coloured versions of the images in the FV14 dataset to the models. The dissimilarity between each item was computed as 1 - Pearson's $r$ to generate the RDMs. For the THINGS dataset, the average features from the 12 images for each concept were used to generate the RDM.

ROI definition

The VOTC was defined by combining functionally defined regions (activation from hearing participants viewing pictures relative to baseline in the OPN95 dataset) and anatomical parcels from the Harvard–Oxford Atlas (probability > 0.2), specifically the posterior and temporooccipital divisions of the inferior temporal gyrus (15#, 16#), the inferior division of the lateral occipital cortex (23#), the posterior division of the parahippocampal gyrus (35#), the lingual gyrus (36#), the posterior division of the temporal fusiform cortex (38#), the temporal occipital fusiform cortex (39#), the occipital fusiform gyrus (40#), the supracalcarine cortex (47#), and the occipital pole (48#). This definition resulted in the selection of 2467 voxels in the left hemisphere and 2420 voxels in the right hemisphere.

Representation similarity analysis

To identify the areas of the VOTC representing the sentence description effect and verbal categorization effect, we conducted RSA using a searchlight procedure (Kriegeskorte et al., 2006) within the defined VOTC mask for each participant.

For the OPN95, SPN95, and FV14 datasets, the $t$ value (corresponding to each object relative to the baseline) images of each item were used to calculate the neural RDMs, and for the THINGS dataset, we used beta-value images instead due to the high signal-to-noise ratio in this dataset. For each voxel within the VOTC mask, multivariate activation patterns within a sphere (radius = 10 mm) centred at that voxel were extracted. Neural RDMs were computed with the Pearson distance within the searchlight sphere. Then, Spearman's rank correlation coefficients between the neural RDM and model-derived visual RDMs was computed, controlling for the effects of models with lower levels of language involvement to determine the *sentence description effect* (correlations with the visual RDMs derived from CLIPvision while controlling for the visual RDMs derived from ResNet-50 and MoCo) and the *verbal categorization effect* (correlations with the visual RDMs derived from ResNet-50 while controlling for the visual RDMs derived from MoCo). Correlation maps were obtained for each participant by moving the searchlight centre across the VOTC mask. These maps were Fisher z-transformed and spatially smoothed with a 6 mm full-width half-maximum (FWHM) Gaussian kernel. The correlation maps were compared to 0 with one-tailed one-sample t tests.

For ROI analysis, multivariate activity patterns for each stimulus within the ROI mask were extracted. Neural RDMs were computed on the basis of Pearson distances and then correlated with the RDM generated by CLIPvision while controlling for the RDMs generated by ResNet-50 and MoCo. The resulting correlation coefficients between the neural and model RDMs were Fisher z-transformed and compared to zero with one-tailed one-sample t tests.

For the comparison between OPN95 and SPN95, Bayesian independent samples t tests were conducted with the Pingouin package (Vallat, 2018) on the mean correlation values within the ROI, with a default Cauchy prior width of r = 0.707 for the effect size on the alternative hypothesis (H1: Hearing $\neq$ Deaf).

Computation of the laterality index

Functional lateralization for each dataset was assessed with the LI via the bootstrap method in the LI tool (Wilke & Lidzba, 2007) for SPM12. For the group-level analysis, the *t* map of the group RSA-derived rho values (Fisher z-transformed, then spatially smoothed using a 6 mm FWHM Gaussian kernel) versus zero in each dataset was entered as inputs to calculate the LI. For the individual-level analysis, the RSA-derived *r* value map of each subject was entered as the input to calculate the LI and compared to 0 using a two-tailed one-sample t test. The Cohen's *d* effect sizes were also computed.

The LI was calculated via the bootstrap method with the following options: the bilateral VOTC mask defined above as an inclusive mask, no exclusive mask, and the default bootstrapping parameters. This method involved the computation of 20 thresholds with equal step lengths ranging from 0 to the maximum *t* value. At each threshold, 100 bootstrapped samples were drawn from both the left and right ROIs, resulting in a total of 200 samples. From these samples, all 10,000 potential LI combinations (100 samples from the left ROIs multiplied by 100 samples from the right ROIs) were calculated for the surviving voxels with the formula [(L − R)/(L + R)]. To mitigate the influence of statistical outliers, only the central 50% of the data were retained and averaged. The group-level LI index ranged from -1 to 1, where a value of -1 indicates complete right lateralization, a value of 1 indicates complete left lateralization, and values between -0.2 and 0.2 are classified as bilateralization (Seghier, 2008).

Study 2: Model-fitting-brain effects tested with lesion models

MRI dataset of patients

*Participants*. Thirty-three stroke patients, demographically matched with the participants in the FV14 database, were recruited following the same task procedure used for the FV14 database (i.e., colour retrieval task). The inclusion criteria for the stroke patients were as follows: aged 20–65 years; right-handedness before stroke onset; normal or corrected-to-normal vision; at least 3 months after stroke;; first symptomatic stroke of ischaemic or intraparenchymal haemorrhagic aetiology; lesion location involving the cortex and/or subcortical WM; no other neurological or psychiatric diseases; ability to perform simple cognitive tasks and understand instructions; and intact object perception. Both the stroke group and the healthy group participated in all the following scans.

*Image acquisition.* Functional and anatomical MR images were collected at the Department of Magnetic Resonance Imaging, First Hospital of Shanxi Medical University, with a 3T Siemens Magnetom Skyra scanner. The scans included task-fMRI, HARDI, high-resolution 3D T1-weighted imaging, 3D T2-weighted imaging and 3D fluid-attenuated inversion-recovery (FLAIR) T2-weighted imaging. Functional images were acquired with a multiband EPI sequence (axial slices, TR = 2000 ms, TE = 30 ms, flip angle = 90°, matrix size = 72 × 72, voxel size = 2.5 × 2.5 × 2.5 mm3, FOV = 180 mm × 180 mm, multiband factor = 2). HARDI images were acquired with a multiband EPI sequence (axial slices, TR = 3000 ms, TE = 100 ms, flip angle = 90°, matrix size = 112 × 112, voxel size = 2 × 2 × 2 mm³, FOV =

224 × 224 mm², multiband factor = 2, diffusion gradient value b set to 0, 1000, and 2000 s/mm², b0 repeated 10 times, and 64 directions each for b1000 and b2000). 3D T1-weighted images were acquired with an MPRAGE sequence (sagittal slices, TR = 2530 ms, TE = 2.88 ms, flip angle = 7°, matrix size = 224 × 256, interpolated to 448 × 512, voxel size = 0.5 × 0.5 × 1 mm³, FOV = 224 × 256 mm²). T2-weighted images were acquired with a Sampling Perfection with Application optimized Contrast using different flip angle Evolution (SPACE) sequence (sagittal slices, TR = 3200 ms, TE = 408 ms, flip angle = 120°, matrix size = 256 × 256, voxel size = 0.9 × 0.9 × 0.9 mm³, FOV = 230 × 230 mm²). FLAIR T2-weighted images were acquired with a FLAIR sequence (sagittal slices, TR = 5000 ms, TE = 394 ms, flip angle = 120°, matrix size = 256 × 256, interpolated to 512 × 512, voxel size = 0.5 × 0.5 × 1 mm³, FOV = 250 × 250 mm²).

*Image analysis.* The functional images were preprocessed and analysed as described for Study 1 above. One or two runs from four stroke patients demonstrated excessive head motion (> 3 mm/3°) and runs were subsequently excluded from the analysis. For the HARDI data, preprocessing was performed using FSL (version 6.07) from Oxford University, including (1) Eddycorrect for correcting distortions and head movement; (2) BET for skull removal; and (3) DTIFIT for building diffusion tensor models and calculating FA maps. FA images were registered to the T1 images with FLIRT and then to MNI space with T1-to-MNI transformation, achieving a voxel size of 2 × 2 × 2 mm³. The derived transformation parameters were used to warp the ROI from MNI space to the native diffusion space via nearest-neighbour interpolation. For the structural MRI data, lesions were drawn by a radiology resident and reviewed by a radiologist using ITK-SNAP. The protocol included (1) coregistration of the T2-weighted and FLAIR images to the 3D T1-weighted images; (2) lesion delineation on the axial FLAIR images, including glioses; (3) lesion image registration to MNI space via normalization parameters from fMRI preprocessing, with a voxel size of 1 × 1 × 1 mm³; and (4) application of a 3 mm smoothing kernel to obtain the brain injury map.

*Defining the WM mask of interest in healthy controls*

WM connections of the healthy individuals were mapped with probabilistic tractography. This technique examines the probability distributions of the fibre orientations in the brain, allowing representation of uncertainty and the delineation of crossing fibres. For datasets with multiple b values, this approach enhances fibre orientation accuracy in the WM and regions near the cortex.

*ROI selection.* We selected the left language-specific regions as defined by Fedorenko et al. (2010) and the left VOTC mask (defined in the ROI definition section). For the controls, the right-hemisphere homologues of the left-hemisphere language regions were also included. All the ROIs were transformed into the diffusion native space for each participant for probabilistic tractography.

*Probabilistic tractography.* Using the HARDI data from the healthy controls with FMRIB's Diffusion Toolbox in FSL, we performed tractography between each pair of ROIs. This procedure focused on (1) connections between the left VOTC mask and each left language region and (2) connections between the left VOTC mask and the right hemisphere homologue of each left-hemisphere language region. Fibre tracking was performed using FSL's BEDPOSTX with default parameters, modelling WM fibre orientations and crossing fibres. Fibre tracking was initiated in both directions, with 5,000 streamlines drawn from each voxel in the ROI with PROBTRACKX 2.0. A cerebrospinal fluid (CSF) mask was used as the exclusion mask. The connectivity distributions were normalized to MNI space and standardized. At the individual participant level, path images were thresholded at a value of 0.1 to remove low-connectivity probability voxels. At the group level, fibre projections present in more than 50% of the individuals within the explicit WM mask (probability > 0.4 in the WM probability map (Fonov et al., 2011)) were retained for analysis.

Correlation analysis of the WM-vision-model effects

To assess the associations between left VOTC-language WM connection integrity and the effect patterns of the different vision models, we performed multiple regression analyses. In these analyses, RSA-derived rho values (Fisher z-transformed) of CLIPvision, ResNet-50 and MoCo were used as predictors of the WM connection integrity values between the left VOTC mask and each language region, with the total lesion volume as a covariate. WM integrity was quantified as the mean FA value within the WM tract mask for each patient. Model significance was measured with F tests and t tests.

To confirm whether the sentence description effect observed in the left VOTC was related to language, partial Spearman's rank correlation analyses were subsequently conducted, with total lesion volume as a covariate, to analyse the relationship between the average integrity of the WM tracts (as measured by the mean FA values) of interest and the sentence description effect within the left VOTC mask. One-sample t tests were performed on the Fisher z-transformed correlation coefficients.

To rule out the possibility that the sentence description effect in the left VOTC was related to other high-order cognitive control functions rather than language, the analyses conducted for the left language regions were also performed on their right-hemisphere homologues.

<u>Brain visualization</u>

The brain results were projected onto the MNI brain surface for visualization with BrainNet Viewer (version 1.7; https://www.nitrc.org/projects/bnv/; RRID: SCR_009446; Xia et al., 2013) with the default 'interpolated' mapping algorithm, unless stated explicitly otherwise.

**Figure**

**a**

| Dataset ID | Reference | Stimulus type | Object domain | Task | Subject (N) |
|---|---|---|---|---|---|
| OPN95 | Wu et al., 2020 | Isolated objects | Animal; Large nonmanipulable object; Small manipulable object | Picture naming by oral language | Healthy, hearing (26) |
| SPN95 | Unpublished | Isolated objects | Animal; Large nonmanipulable object; Small manipulable object | Picture naming by sign language | Healthy, early deaf (29) |
| FV14 | Unpublished | Isolated objects | Fruits; Vegetables | Object Color Knowledge Judgment | Healthy, hearing (33) |
| THINGS | Hebart et al., 2023 | Naturalistic | Animal; Large nonmanipulable object; Small manipulable object; Foods; | Oddball detection | Healthy, hearing (3) |

**b**

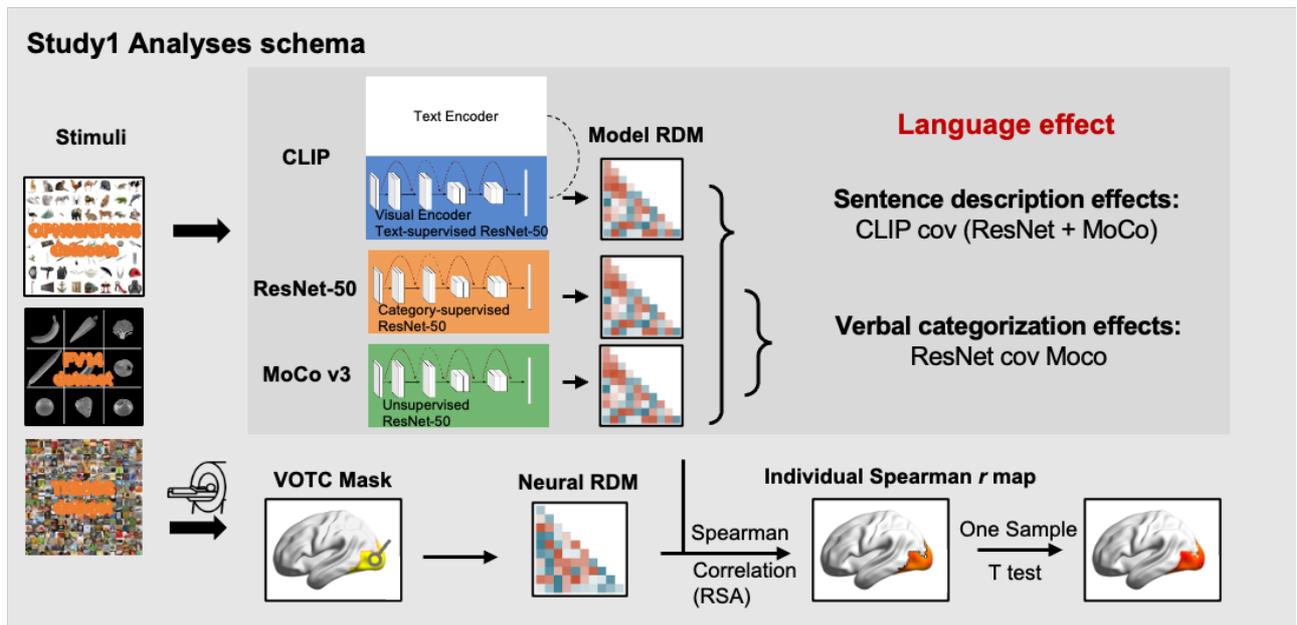

**Fig. 1.** Overview of the fMRI datasets, vision models, and Study 1 analysis schema. **a.** Table listing the four fMRI datasets (OPN95, SPN95, FV14, and THINGS) used in the present study's analyses and detailing the references, stimulus types, object domains, tasks, and participant numbers. Note that the participants in the FV14 dataset served as healthy controls to the patient group in Study 2. **b.** The analysis pipeline for Study 1. We compared three vision models in terms of their fits to the neural responses in the ventral occipitotemporal cortex (VOTC): the visual encoder of OpenAI's CLIP (Radford et al., 2021; hereinafter referred to as CLIP), ResNet-50 (He et al., 2015), and MoCo v3 (Chen et al., 2021). All three models share the same ResNet-50 backbone but differ in their training supervision: natural language texts for CLIPvision, human-generated category labels for ResNet-50 and self-supervised learning for MoCo v3. A set of object stimuli spanning multiple categories was presented to both the vision models and human participants during fMRI data collection. A searchlight representational similarity analysis (RSA) approach was then applied to compare neural responses with vision model responses. Specifically, for each voxel in the VOTC mask, a spherical ROI was created to extract activation patterns for each object, on the basis of which Pearson's distance of the neural responses for each object pair was computed to construct the neural representational dissimilarity matrices (RDMs). Spearman's partial correlation coefficients were computed between the neural and model-based RDMs (from CLIPvision, ResNet-50, and MoCo v3), which resulted in a set of Fisher z-transformed *r* VOTC maps for each subject. A one-sample t test across participants was used to identify regions that significantly tracked each model's representational structure, enabling an assessment of language-related effects.

Here, the "*sentence description*" effect is captured as the partial correlation coefficient between the CLIP RDM and the neural RDMs, controlling for the ResNet- and MoCo-derived RDMs, whereas the "*verbal categorization*" effect is captured by the partial correlation coefficient between the ResNet RDM and the neural RDMs, controlling for the MoCo RDM. See "Methods" for details on the preprocessing steps, ROI definitions, and statistical analyses.

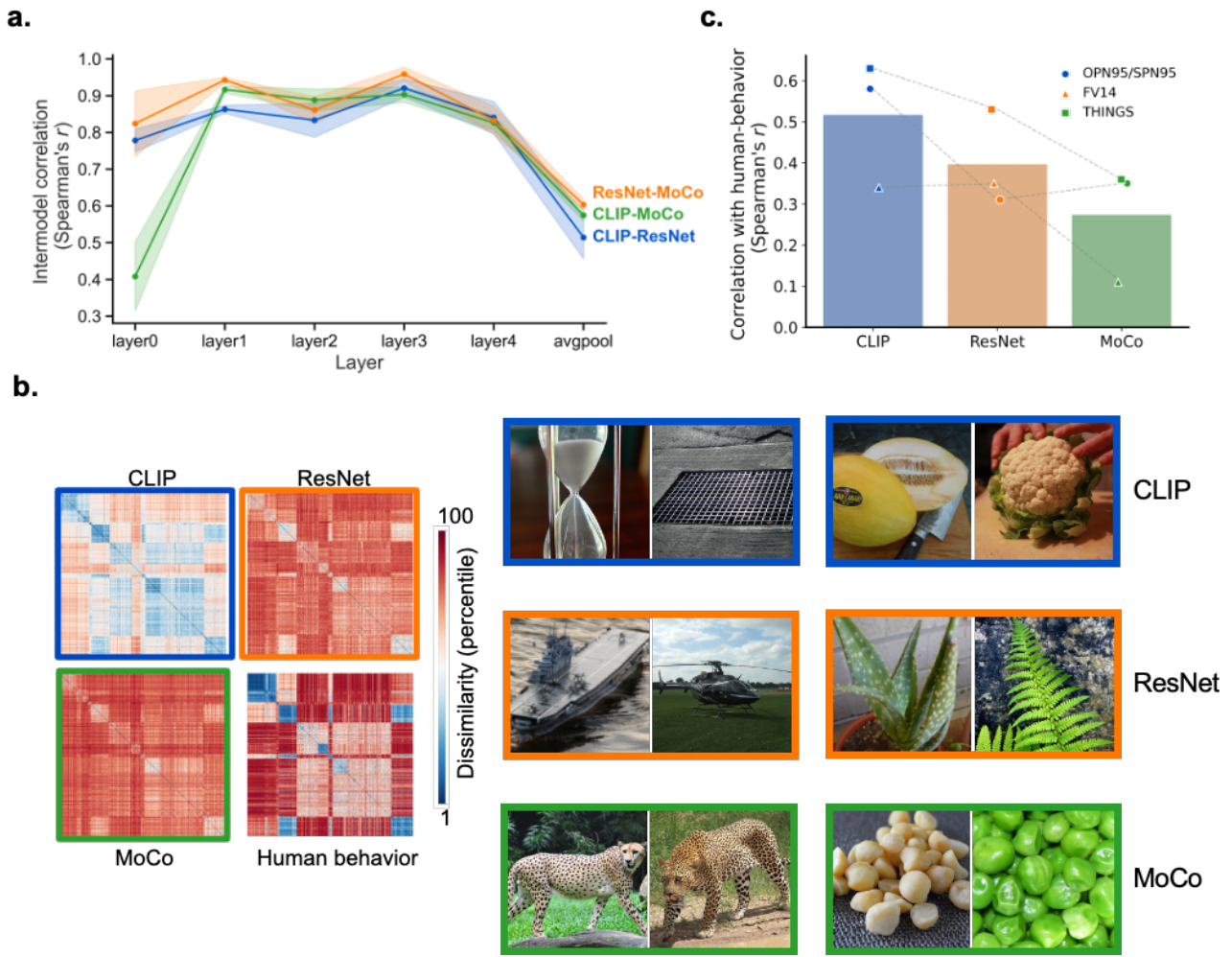

**Fig. 2.** Vision model RDM intercorrelations and visualization. **a.** Intermodel Spearman's correlation between the representational dissimilarity matrices (RDMs) at each layer of CLIP (visual encoder), ResNet-50, and MoCo v3 averaged over the datasets used in this study (OPN95/SPN95, FV14, THINGS). Each point represents the mean correlation coefficient, and the shaded regions denote the standard error across the datasets. The average pooling layer ("avgpool") yielded the lowest intermodel correlation coefficients and was therefore selected for the subsequent analyses. **b.** The left panel shows the RDMs from the 3 vision models for the THINGS dataset, where red corresponds to the least similar object pairs and blue corresponds to the most similar object pairs. A human-behaviour object-relatedness RDM derived with a three-alternative odd-one-out paradigm is also shown (Hebart et al. 2023). The right panel shows the top 2 representative image pairs with the highest similarity for each model, computed after partialing out the RDMs of the other two models, illustrating how these models group object concepts differently. For the THINGS dataset, each object concept is represented by 12 images, across which the model-derived features were averaged to construct the concept-level RDM (see "Methods" for details). **c.** Spearman correlation coefficients between each model's RDM and the human-behaviour object-relatedness RDM for all the datasets used in this study (behavioural RDMs were derived via pairwise ratings for OPN95/SPN95 and FV14). The bar heights represent the mean correlation coefficient across the datasets, whereas the overlaid scatter points represent individual dataset values. Dotted lines connect points from the same dataset, highlighting the variability in alignment between the models and human behaviour.

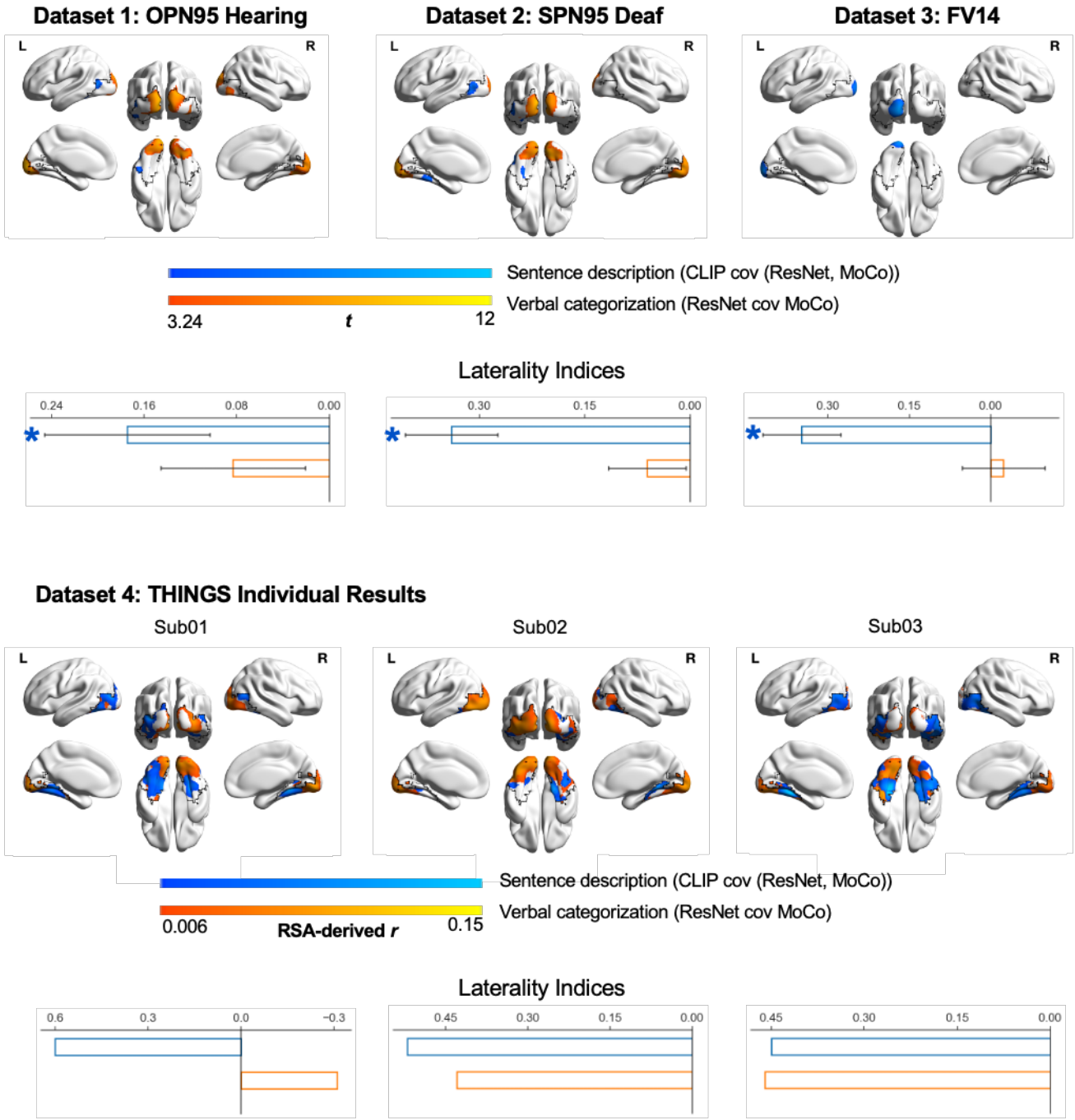

**Fig. 3.** Searchlight RSA brain maps for the sentence description effect and verbal categorization effect in the ventral occipitotemporal cortex (VOTC) and their literalization effects. The black contour lines on each brain map mark the boundary of the VOTC mask used for the searchlight analyses. The top row (Datasets 1–3) shows the group-level maps for OPN95 (hearing participants), SPN95 (deaf participants), and FV14 (healthy controls in Study 2), with colour bars indicating *t* values (blue, the sentence description effect, defined as the partial correlation coefficient between the CLIP RDM and the neural RDMs after controlling for the ResNet and MoCo RDMs; orange, the verbal categorization effect, defined as the partial correlation coefficient between the ResNet RDM and the neural RDMs after controlling for the MoCo RDM). The bar plots below illustrate the laterality indices (LIs) of the two effects of interest in the VOTC; bar heights represent the group mean, and error bars represent the standard error (SE) across participants. Asterisks indicate that LIs are significantly greater than zero. The bottom row (Dataset 4) represents the individual-level maps from three participants (Sub01, Sub02, Sub03) in the THINGS dataset, with colour bars reflecting Spearman's correlation coefficient ($r$). The bar plots below depict the LIs at the single-participant level. All the brain maps were thresholded at voxel-level $p < 0.001$, one-tailed, cluster-level familywise error (FWE)-corrected $p < 0.05$.

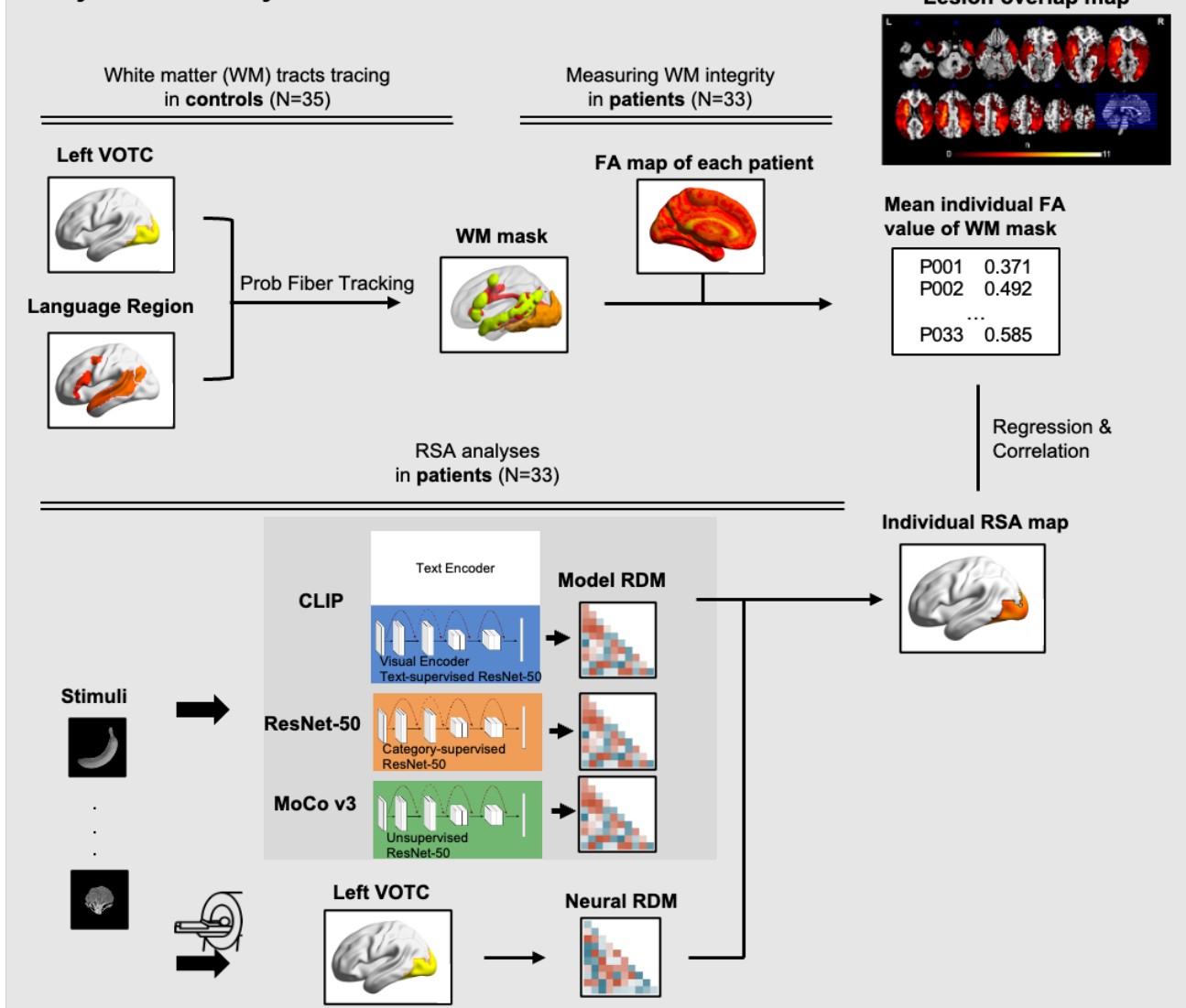

**Fig. 4.** Study 2 analysis workflow linking white matter (WM) integrity and vision model-based neural representations in patients with chronic stroke. The top panel shows the workflow for quantifying WM integrity in patients. Probabilistic tractography was first performed in the native spaces of the healthy controls (N = 35; the same participants in the FV14 dataset of Study 1), taking six regions of interest (ROIs) within the language network (along with the overarching language mask) (Fedorenko et al., 2010) as the "seed" points and the left ventral occipitotemporal cortex (VOTC) as the "target." Each tractography-derived map was normalized, scaled to its maximum voxel intensity, binarized at 0.1, and then aggregated across controls to yield the group-level WM mask (within an explicit WM template; probability > 0.4). In the patient group (N = 33), we quantified WM integrity between language regions and the left VOTC by extracting the mean fractional anisotropy (FA) within each group-level WM mask. The bottom panel shows the RSA analysis workflow of the task fMRI data in these patients following a procedure similar to that of Study 1 but performed at the ROI level. Neural representational dissimilarity matrices (RDMs) were generated on the basis of the activation patterns of the left VOTC, and each patient's model–brain similarity was computed by separately correlating (Spearman's correlation) the neural RDMs with the model-derived RDMs of CLIP, ResNet-50 and MoCo v3. Finally, we performed multiple linear regression and correlation analyses across the patient group, relating WM integrity (i.e., FA values) in specific tracts to each patient's RSA values in the left VOTC. The upper right panel shows the lesion distribution map of the 33 patients, in which the n value of each voxel represents the number of patients with lesions at that voxel.

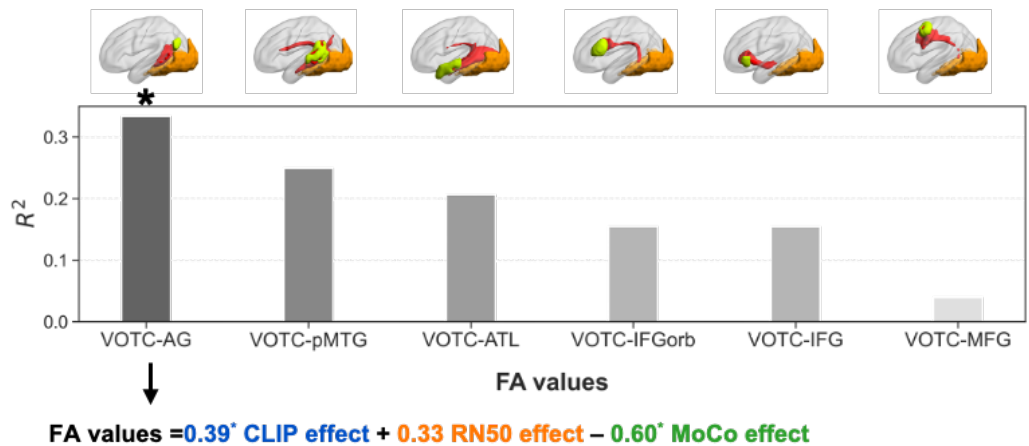

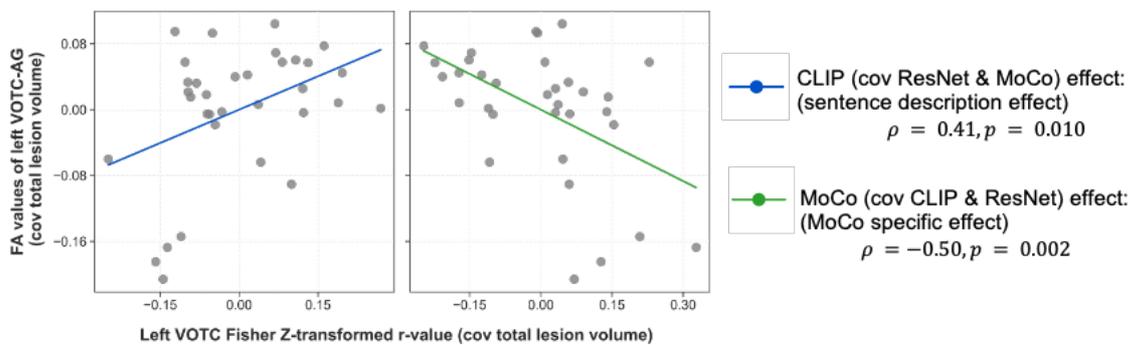

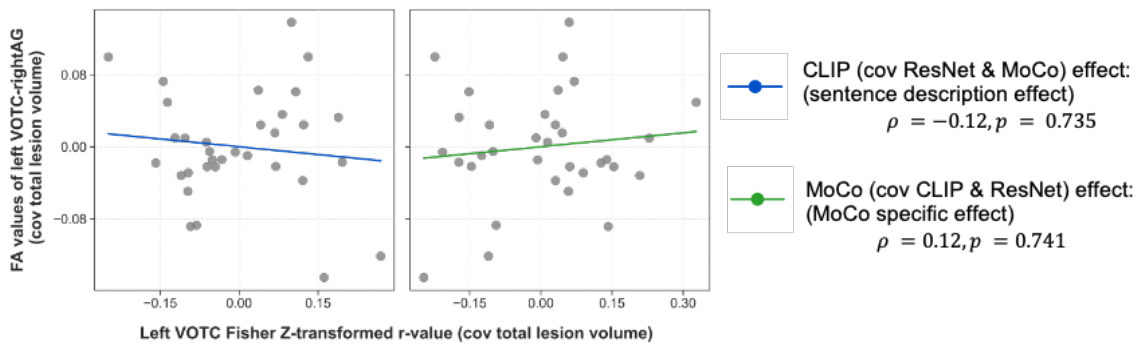

**Fig. 5.** Relationships between left VOTC–language white matter (WM) tract integrity and model-based neural representations in the patient group. **a.** Multiple regression results. The bars represent the proportion of variance ($R^2$) explained by the RSA-derived Spearman's correlation coefficient for each patient (CLIP, ResNet-50, and MoCo), as well as total lesion volume, when predicting the FA values for each left VOTC–language tract. The asterisk (*) on the left VOTC–AG bar indicates that the overall regression model is significant (F test). The regression equation below the bar chart shows the standardized regression coefficients for the three model-based effects (CLIP, ResNet-50, MoCo) in predicting the FA values of the left VOTC–AG tract (intercept omitted). The CLIP and MoCo coefficients reached statistical significance, as denoted by asterisks. **b.** Partial Spearman's correlation analyses focusing on the relationship between left VOTC–AG tract integrity and vision model-based neural representations in the left VOTC. Each scatter point represents one patient, and the best-fit line is derived from the Spearman correlation. The sentence description effect (correlation coefficient between the CLIP RDM and the left VOTC neural RDM, while partialing out the ResNet and MoCo RDMs) and the MoCo-specific effect (correlation coefficient between the MoCo RDM and the left VOTC neural RDM, while partialing out the CLIP and ResNet RDMs) are each correlated with FA values in the left VOTC–AG tract, controlling for total lesion volume. One-tailed *p* values are reported. **c.** Validation analysis for the nonlinguistic contributions of the AG, using the left VOTC–right AG tract. Each scatter point represents an individual patient, with the best-fit line determined by Spearman's correlation. The same partial correlation approach (controlling for total lesion volume) was applied. No significant relationships were observed, highlighting the specificity of the left-lateralized VOTC–AG pathway for these model-based effects.

Table 1. Results of Regression Analysis: Impact of CLIP, ResNet, and MoCo on FA values of left VOTC-language tracts

| FA values of WM Tract (dependent variable) | $R^2$ | F stats | Standardized Beta | | |
|---|---|---|---|---|---|
| | | | CLIP | ResNet | MoCo |
| VOTC-Lang | 0.20 | 1.76 | 0.19 | 0.21 | -0.44* |
| VOTC-AG | 0.33 | **3.50*** | **0.38*** | 0.32# | **-0.58*** |
| VOTC-ATL | 0.21 | 1.82 | 0.31 | 0.26 | -0.47* |
| VOTC-IFG | 0.15 | 1.28 | -0.03 | 0.09 | -0.27 |
| VOTC-IFGorb | 0.15 | 1.28 | -0.10 | 0.01 | -0.22 |
| VOTC-MFG | 0.04 | 0.28 | -0.04 | -0.04 | -0.13 |
| VOTC-pMTG | 0.24 | 2.32 | 0.31 | 0.28 | -0.49* |

All models have the same degrees of freedom (df1 = 4, df2 = 28). Significance: *$p < 0.05$, **$p < 0.01$, two-tailed. *Abbreviations: FA, fractional anisotropy; WM, white matter; VOTC, (left) ventral occpitotemporal cortex; AG, (left) angular gyrus; ATL, (left) anterior temporal lobe; IFG, (left) inferior frontal gyrus; IFGorb, (left) orbital part; MFG, (left) middle frontal gyrus; pMTG, (left) posterior middle temporal gyrus.*

**Table 1.** Multiple regression results predicting white matter (WM) integrity in the left ventral occipitotemporal cortex (VOTC)–language tracts, with vision model-based neural representations used as predictors (CLIP, ResNet-50, and MoCo) besides total lesion volume.

**Supplementary Captions**

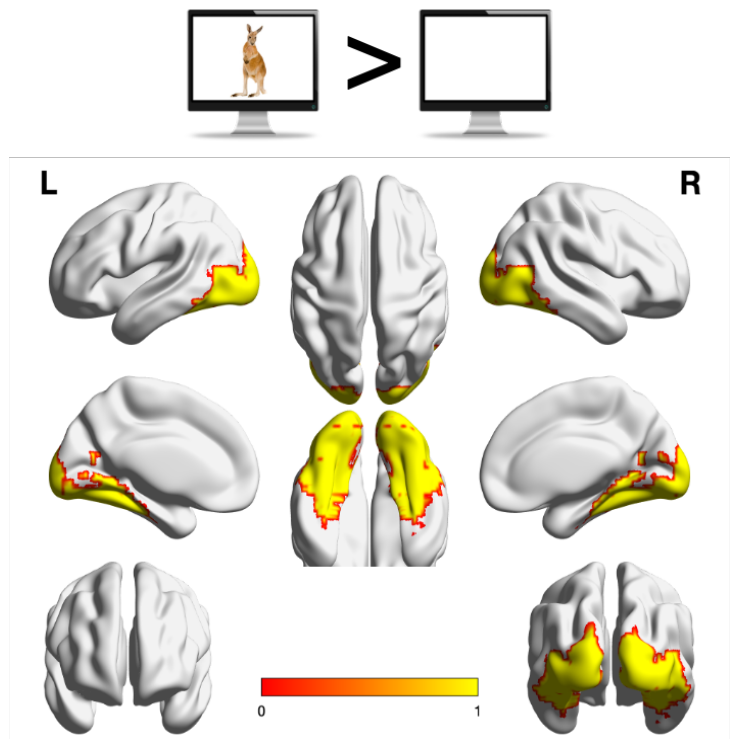

**Supplementary Fig. 1.** The ventral occipitotemporal cortex (VOTC) mask used in the analyses, defined on the basis of regions showing stronger activation to all pictures relative to baseline in hearing participants from the OPN95 dataset ($q < 0.05$, FDR-corrected). The functional mask was further constrained by the following anatomical parcels (Harvard–Oxford Atlas, probability > 0.2): the posterior and temporooccipital divisions of the inferior temporal gyrus (#15, #16), the inferior division of the lateral occipital cortex (#23), the posterior division of the parahippocampal gyrus (#35), the lingual gyrus (#36), the posterior division of the temporal fusiform cortex (#38), the temporal occipital fusiform cortex (#39), the occipital fusiform gyrus (#40), the supracalcarine cortex (#47), and the occipital pole (#48).